\documentclass[preprint,showpacs,floatfix,superscriptaddress]{revtex4}

\usepackage{courier}
\usepackage{times}
\usepackage{amsmath}
\usepackage{amssymb}
\usepackage{graphicx} 

\begin{document}
\title{Bosonic Helium droplets with cationic impurities: onset
of electrostriction and snowball effects from quantum calculations}

\author{E. Coccia}
\affiliation{Department of Chemistry, University of Rome ``La
Sapienza''and CNISM, Piazzale  A. Moro 5, 00185 Rome, Italy}
\author{E. Bodo}
\affiliation{Department of Chemistry, University of Rome ``La
Sapienza''and CNISM, Piazzale  A. Moro 5, 00185 Rome, Italy}
\author{F. Marinetti}
\affiliation{Department of Chemistry, University of Rome ``La
Sapienza''and CNISM, Piazzale  A. Moro 5, 00185 Rome, Italy}
\author{F. A. Gianturco\footnote{corresponding author. e-mail:
      fa.gianturco@caspur.it  fax: +39-06-49913305}}
\affiliation{Department of Chemistry, University of Rome ``La
Sapienza''and CNISM, Piazzale  A. Moro 5, 00185 Rome, Italy}
\author{E. Yildrim}
\affiliation{Istanbul Technical University, Chemistry Department, 80626
Maslak, Istanbul, Turkey}
\author{M. Yurtsever}
\affiliation{Istanbul Technical University, Chemistry Department, 80626
Maslak, Istanbul, Turkey}
\author{E. Yurtsever}  
\affiliation{Chem. Dep., Ko\c c University Istanbul 34450 Turkey}

\begin{abstract}
Variational MonteCarlo and Diffusion MonteCarlo calculations have been carried
out for cations like Li$^+$, Na$^+$ and K$^+$ as dopants of small helium
clusters over a range of cluster sizes up to about 12 solvent atoms. The
interaction has been modelled through a sum-of-potential picture that
disregards higher order effects beyond atom-atom and atom-ion contributions.
The latter were obtained from highly correlated ab-initio calculations over a
broad range of interatomic distances.

This study focuses on two of the most striking features of the microsolvation
in a quantum solvent of a cationic dopant: electrostriction and snowball
effects. They are here discussed in detail and in relation with the
nanoscopic properties of the interaction forces at play within a fully quantum
picture of the clusters features. 

\end{abstract}


\maketitle

%----------------------------------------------------------------------------
\section{Introduction}
Our knowledge and understanding of He droplets has witnessed a tremendous
growth in recent years, a fact which has contributed, to a large extent, to
keep the physics and the chemical physics of liquid helium, and of quantum
fluids in general, a lively research field attracting the attention of
researchers from many-body theory, molecular and atomic spectroscopy and
quantum molecular methods
\cite{cl:toennies98,cl:toennies01,cl:stienkemeier01,cl:toennies04,
cl:stienkemeier06,cl:barranco06}.
From the standpoint of picking-up possible impurities, the He droplets are able
to do so with any species with which they collide \cite{cl:toennies04} and the
latter can then reside either in the inside of the droplet or at its surface
depending on the strength of the helium-dopant interaction with respect to the
He-He one (e.g. see refs. \cite{zb:bodo04,zb:bodo05-e}).

One question about $^4$He clusters frequently discussed in the literature is
the possible existence of magic numbers which are  associated with
particular configurations that are energetically more stable than neighboring
ones and contain a specific number of adatoms \cite{cl:guardiola06}. The
theoretical estimates for He droplets, in fact, suggested that in the pure
aggregates well defined structures are not expected \cite{cl:melzer84} and
even recent calculations and experiments \cite{cl:guardiola06} connected the
presence of such ``magic numbers'' observed in $^4$He clusters of small size to
surface excitation processes rather than to special stability effects. 

The situation, however, is drastically modified when a cationic impurity  is
being picked up by the droplet: the charged dopant is expected to modify the
local environment of the liquid and to give rise to a region of increased
density and of electrostriction effects which can cause regular structures to
arise around the impurity (snowball effect) \cite{cl:atkins59,cl:takahashi00}.
Earlier theoretical studies using ``shadow'' wavefunctions
\cite{cl:macfarland94} and modelling large clusters with 60-70 
He atoms, have predicted the formation of an easily identifiable first
solvation shell \cite{cl:galli01} where the helium density is well above the
critical freezing density of bulk helium so that the first shell can be
characterized by a solid-like structure and order parameter \cite{cl:rossi04}.
The snowball structure and composition, however, in contrast to the
simple snowball
model, seem to depend on the nature of the central ion. For the three alkali
ions that we are going to discuss in the present work, Galli et al.
\cite{cl:galli01} had initially reported a
snowball of 8 atoms for Li$^+$, of 10 for Na$^+$ and 12 for K$^+$. A more recent
study by the same group \cite{cl:rossi04} has improved their result to 12 for
Na$^+$ and 15 for K$^+$. A Path Integral Monte Carlo calculation has
also been performed on Na$^+$(He)$_{100}$ and a triple layer
structure was found \cite{cl:nakayama00}: the first shell was found solid-like
while the third one was like liquid helium. Furthermore, our previous DMC
calculations 
\cite{zb:bodo05-f,zb:bodo05-e} have found a first solvation shell of 10 helium
atoms for Li$^+$ which is in qualitative agreement with a spatial
configuration corresponding to the antiprism classical structure.

The purpose of the present work is to show that quantum stochastic calculations
in small $^4$He clusters that contain alkali cations can provide
detailed evidence for
both ``crowding'' of the solvent around the impurity (electrostriction) and the
appearance of regular structures of largely equivalent adatoms (snowballs).
Contrary to what has been previously done on the  clusters doped with alkali
metal ions we shall focus our attention only on the smaller clusters and we
shall follow the evolution of their geometrical and energetic properties
along the series from the dimers to the moieties containing 10-12 helium atoms. 
It is worth pointing out that the Potential Energy Curves (PECs) we use in the
present work are
different from those used in the previously mentioned studies which
were instead taken from the model potentials of ref. \cite{cl:koutselos90}. In
general our fully ab-initio PECs are slightly deeper (5\%) and our $r_{eq}$
values are by a few percent smaller.  Our new PECs are described in the next
Section, while Section I describes our computational methods and
reports also to the features of the diatoms. In Section
\ref{trimers} we will present the results of the triatomic clusters and Section
\ref{clu} will describe our results for the larger clusters and their
snowballs structures. Finally we will summarize our present conclusions in
Section \ref{last}.

%----------------------------------------------------------------------------
\section{The Potentials}
The overall interaction in the cluster is modelled using the sum-of-potential
approximation. For a cluster of $n$ helium atoms containing only one ionic
impurity we therefore have:
\begin{equation}
 V_{\mathrm{TOT}}=\sum_{i=1}^n V_{\mathrm{M^++He}} +
\sum_{i<j}V_{\mathrm{He-He}}\label{pot}
\end{equation}
We have taken the He-He potential from Tang et al.
\cite{cl:tang95} while for the M$^+$-He potential energy curves
(PECs) we
have used accurate potentials calculated by us. The Li$^+$-He potential has been
already presented in refs. \cite{zb:bodo05-e,zb:bodo05-f} while the PECs
for K$^+$-He and  Na$^+$-He are given here for the first time. They have been
calculated using the Coupled-Cluster Singles
and Doubles including triple excitations
non-iteratively CCSD(T) with an aug-cc-pV5Z quality basis set \cite{ab:lee02}.
In the case of K$^+$ the effective core
potential (ECP10MWB) \cite{ab:leininger96} was used. The calculations were
carried out with the MOLPRO software \cite{prg:MOLPRO_brief} and basis
set superposition errors have been corrected via the counterpoise method
\cite{ab:boys70}. 

The three potentials compare very well with various previous calculations
and models present in the literature, in particular we report in Table
\ref{livelli} a comparison of our
results with those obtained recently by  Viehland et al
\cite{cl:viehland03, cl:viehland04} for K$^+$
and Na$^+$ and by Sold\'{a}n et al \cite{ab:soldan01} for Li$^+$. The data
reported in the Table \ref{livelli} are $D_e$ and $R_e$. 
\begin{table}
 \begin{tabular}{c|cc|cc}
\hline
\multicolumn{1}{l}{Specie} \vline & \multicolumn{2}{c}{$R_e$(a.u.)} \vline &
\multicolumn{2}{c}{$D_e$(cm$^-1$)}  \\
\hline
Li$^+$-He  & 1.897$^a$ & 1.898$^d$ & 649.1$^a$ & 646.1$^d$\\
Na$^+$-He & 2.324$^b$ & 2.323$^d$ & 329.1$^b$ & 331.3$^d$ \\
K$^+$-He  & 2.825$^c$ & 2.839$^d$ & 185.4$^c$ & 179.9$^d$ \\
\hline 
\end{tabular}
\caption{\small
$^a$ From ref. \cite{ab:soldan01};
$^b$ From ref. \cite{cl:viehland03}; $^c$ From ref. \cite{cl:viehland04}; 
$^d$ Present results.
}\label{livelli}
\end{table}
A more detailed comparison of recent and less recent data for the potentials can
be found in refs. \cite{ab:soldan01,zb:bodo05-e,cl:viehland03,cl:viehland04}. 

The calculated ab-initio points have been  fitted with an analytical
expression  which included the long range induction tail which has been
chosen in order to yield at large distances the correct charge-He polarization
potential $-\alpha(\mathrm{He})/2r^4$ with $\alpha(\mathrm{He})$=1.38 a.u.. 
The potential functions and parameters are available on requests from the authors.
More details on these potentials and on the  bound states supported by them can be found  in ref. \cite{zb:bodo07}.

When calculating the global interaction in the cluster using the
sum-of-potentials approximation one neglects the contributions
due to many-body effects. While these contributions are very small in neutral
He clusters \cite{cl:baccarelli05}, they may be of some importance when an ionic
impurity is present. In the latter case, in fact, the main contribution to the
many body interaction terms comes from the fact that the He atoms near the
impurity are polarized by the charge and acquire a small dipole moment. These dipoles are
oriented in such a way as to contribute with a small positive interaction
energy. Furthermore, in Ref. \cite{cl:lenzer01} for the analogous situation of
the anionic dopant Cl$^-$ in Ar clusters it has been shown that the
inclusion of the leading terms of such 3-body forces (3B) does not alter
substantially the energetics and the structure of the clusters.
In refs. \cite{zb:bodo05-e,zb:bodo05-f} we have already discussed this effect
in detail by carrying out a comparison of the sum-of-potentials minimum
geometries with fully ab-initio ones. Our findings indicate that the addition of many-body effects does not substantially modify the geometries of the small clusters of interest in the present study.

%----------------------------------------------------------------------------
\section{The Monte Carlo calculations}\label{mc}

Our numerical procedure is based on a two-step calculation. We first optimize
the trial wavefunction using Variational Monte Carlo (VMC)
(see for example ref. \cite{mc:bressanini02} for a general discussion and ref.
\cite{mc:mushinski94} for the specific implementation). The second step is that
of using the optimized wavefunction in a Diffusion Monte Carlo (DMC) algorithm
\cite{mc:anderson95,mc:ceperly86,mc:viel02}  in order to obtain the
energy and the geometric distributions of the cluster. We have discussed our
DMC procedure more specifically in earlier work
\cite{mc:paesani02-a,mc:paesani02-b,cl:paesani03} and will therefore not be
repeating it here. The form of the
triel wavefunction is the familiar Jastrow function where both the He-He  and the
ion-He pairs have the form
\begin{equation}
 \log
\Psi_T(r)=-\left(\frac{p_5}{r^5}+\frac{p_3}{r^3}+\frac{p_2}{r^2}
+p_1r+p_0\log r\right)
\end{equation}
In the VMC step, the 10 parameters of the wavefunctions (He-He and ion-He) are
optimized using a standard Powell algorithm  \cite{bk:numrec-book}
which minimizes a suitable estimator such as the variance of the local energy 
with a constant population.
The population is then updated using
the Metropolis sampling algorithm with a fixed wavefunction. This cycle is
repeated until convergence is achieved. We noticed that in order to get a faster
convergence we had to minimize in the first instance the second moment of the
distribution given by 
\begin{equation}
 \mu^2_{E_R}(H)=\sigma^2(H)+(<H>-E_R)^2
\end{equation}
where $E_R$ is a suitable reference energy near the ground state energy.  In a
second calculation, we then  minimize directly the variance $\sigma^2(H)$ .
The parameters that we obtain are used inside a DMC program that calculates the
energies and the radial distribution functions  as an overall average over the walkers ensemble 
sampled by the mixed distribution $\Psi\cdot\Psi_T$:
\begin{equation}
\left\langle R\right\rangle
\approx\frac{\sum_{i}^{M}w_{i}\mathbf{R}_{i}}{\sum_{i}^{M}w_{i}}.
\label{dmc_integrale}
\end{equation}
In order to describe the cluster ``geometry'', we use different
 distances $\mathbf{R}_i$ such as the ion-He and He-He ones or the distance of
the atoms from the center of mass or from the geometric center (see ref.
\cite{cl:dipaola05} for a more detailed definition and a further discussion on
such quantities).

One  important effect found in our work is the changing ``size'' of the alkali
partner ionic core within each cluster.. The average Ion-He distances in the $n$=1 complexes 
range from $\sim2$
\AA~ in Li$^+$ to $\sim2.5$ \AA~ in Na$^+$ and to $\sim3$ \AA~ in K$^+$. As we
shall see below, the size of the ion and the resulting steric effects on the solvent atom collocations are key
factors in determining the  features and the geometric structures of the first
solvation shell.
On the other hand, the radial delocalization of the He in the $n=1$ complex
remains almost the same along the series: the
average distances and their FWHM are 2.01$\pm$0.40, 2.46$\pm$0.53,
3.01$\pm$0.55 \AA~ for Li$^+$, Na$^+$ and K$^+$ respectively, while the minima
are located at 1.91, 2.30, 2.93 \AA. 

%----------------------------------------------------------------------------
\section{The triatomic systems}\label{trimers}

Since we have negelected the 3B contributions to the total interaction, the trimers are the natural candidate on
which we could estimate the effect of such corrections. As mentioned
above,  the main contribution to the 3B terms in our ionic systems
originates from the dipole-dipole interaction which is due to the repulsive
forces between the polarized helium atoms located near the charge. As the induced
dipole is inversely proportional to the distance from the ion, the size of this
contribution is likely to be of
some importance only for the the Li$^+$ clusters where the helium atoms
are located the nearest to the ionic center.
In the case of lithium, we have already shown
 with fully ab-initio calculations \cite{zb:bodo05-e,zb:bodo05-f} that the
correct inclusion of all the many-body interactions makes the
minimum structure of Li$^+$He$_2$ linear. On the contrary, when using the
sum-of-potential approximation the minimum structure is triangular. 
Although 3B potential terms
seems to play an important role in such small systems, we expect that  it
should become increasingly less important in
larger cluster, where the crowding
of the He atoms around the Li$^+$ core makes
the repulsive short range He-He interaction the driving force in determining the
geometry. Indeed, a classical
minimization of a slightly larger species, the Li$^+$He$_4$ cluster, when done
by either including or not including the 3B effect, produces
essentially the same regular tetrahedral geometry, the only noticeable
difference being the R(Li$^+$-He) distance changing from 1.89 to 1.91 \AA~
and the R(He-He) distance changing from 3.09 to 3.13 \AA~ with 3B forces.

\begin{figure}
 \includegraphics[height=0.9\textheight]{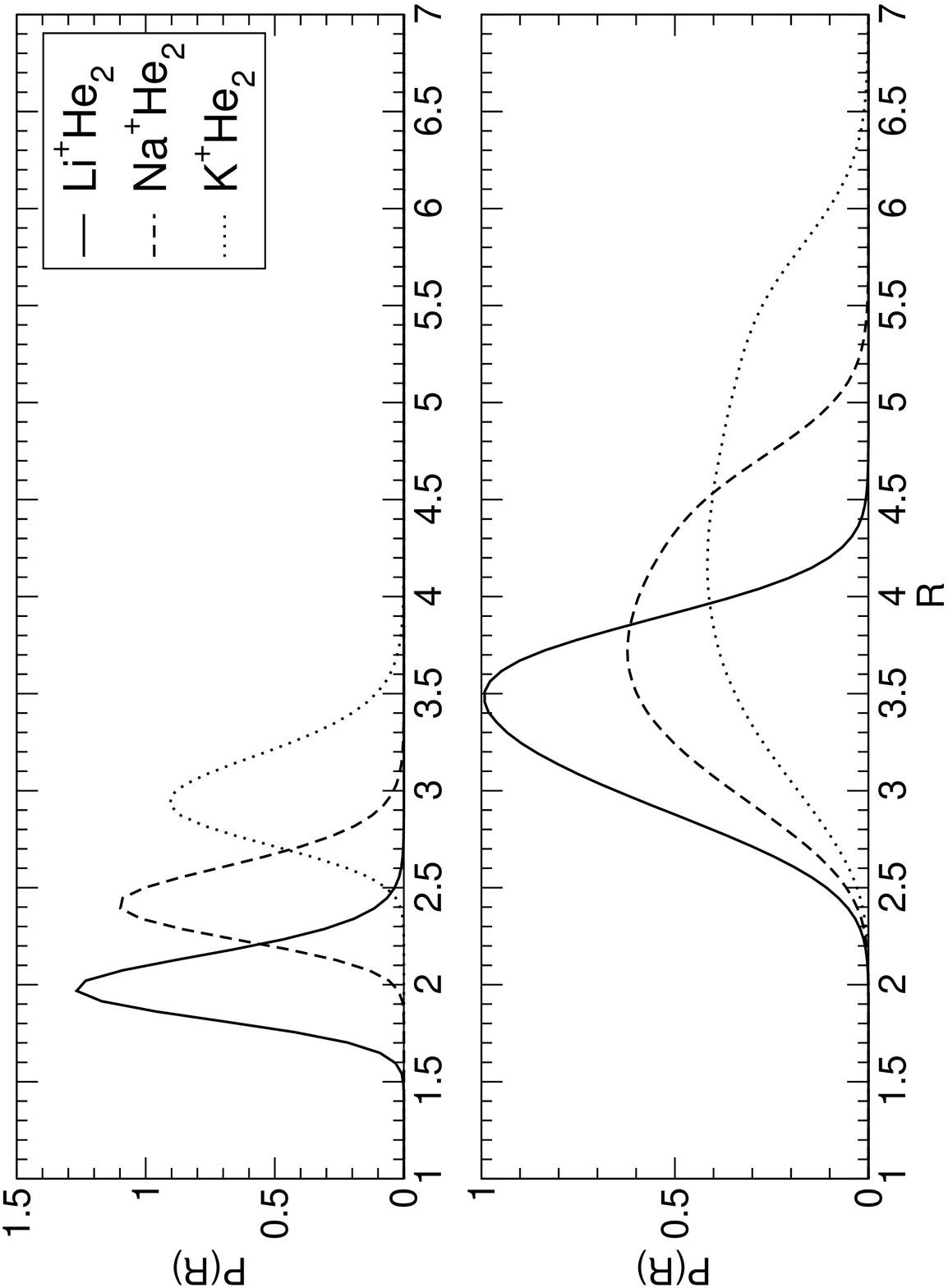}
\label{tri-rad}\caption{}
\end{figure}

We report in
Table \ref{geom} the geometric parameters of
the optimized structures of the MHe$_2^+$ clusters.. The data reported as ``Quantum'' are the
simple averages over the final quantum distributions obtained from the DMC
simulation using the sum-of-potentials approximation. The values labelled as
``classical'' are a simple ``stick-and-ball'' classical optimization that we
obtained using the sum-of-potentials approximation. 
\begin{table}
\begin{tabular}{lccc}
\hline 
\hline
Specie & $^{7}$Li$^{4}$He & $^{23}$Na$^{4}$He &  $^{39}$K$^{4}$He \\
\hline 
\multicolumn{4}{c}{Quantum}\\
$<\theta>$    & 119.9 & 106.1 & 98.6 \\
$<R>$(He-He)  & 3.38  & 3.80    & 4.38 \\
$<R>$(ion-He)   & 2.01  & 2.45  & 2.98 \\
\hline
\multicolumn{4}{c}{Classical}\\
$\theta$      & 102.7 & 79.2  & 63.2 \\ 
R(He-He)      &  2.96 & 2.96  & 2.96\\
R(ion-He)     & 1.90  & 2.32  & 2.83\\
\hline
\hline
\end{tabular}
\caption{Classical geometries and quantum average values for the MHe$_2^+$
Clusters}
\label{geom}
\end{table}
When comparing the classical and quantum geometrical parameters reported in
Table \ref{geom} we see that our DMC results produce quantum distributions whose
averages are in qualitative
agreement with the classical geometries calculated with the sum-of-potentials
approximation. Obviously the quantum values imply a physical picture of 
He atoms which are delocalized in terms of both radial and angular density, 

In figure \ref{tri-rad} we report the radial distributions of the Ion-He
(upper panel) and He-He distances (lower panel).

We can see from the upper panel of Figure
\ref{tri-rad} that the two He atoms are located
around the ion at an average distance of 2.0, 2.5 and
3.0 \AA~ for Li$^+$, Na$^+$ and K$^+$ respectively. 
As one can see in the lower panel of Figure \ref{tri-rad}, the quantum
distribution of the He-He distance is much more delocalized than those of the
Ion-He ones. Nevertheless, the He-He relative motion for the
Li$^+$He$_2$ cluster is still confined to a relatively small region of about
1.5 \AA. This region extends up to 2.5 \AA~ for Na$^+$ while in the case of  K$^+$  the
relative He-He distance extends up to 7 \AA~ on a range of slightly less
than 4.0 \AA. This effect is obviously due to  the weaker potential which exists
for the K$^+$ species with respect to that of
Li$^+$, a feature which therefore induces a more marked delocalization of
the relative motion of the adatoms when one goes from Li$^+$ to K$^+$.
In other words  a  ''triangular`` geometry in the three species is  clearly visible although the quantum
''triangle`` turns out to be increasingly ''floppier`` when going from Li$^+$He$_2$ to the K$^+$He$_2$ moiety.

%----------------------------------------------------------------------------
\section{The A$^+$H\lowercase{e}$_n$ clusters}\label{clu}

\subsection{The Energy Landscape}
\begin{figure}
 \includegraphics[height=0.9\textheight]{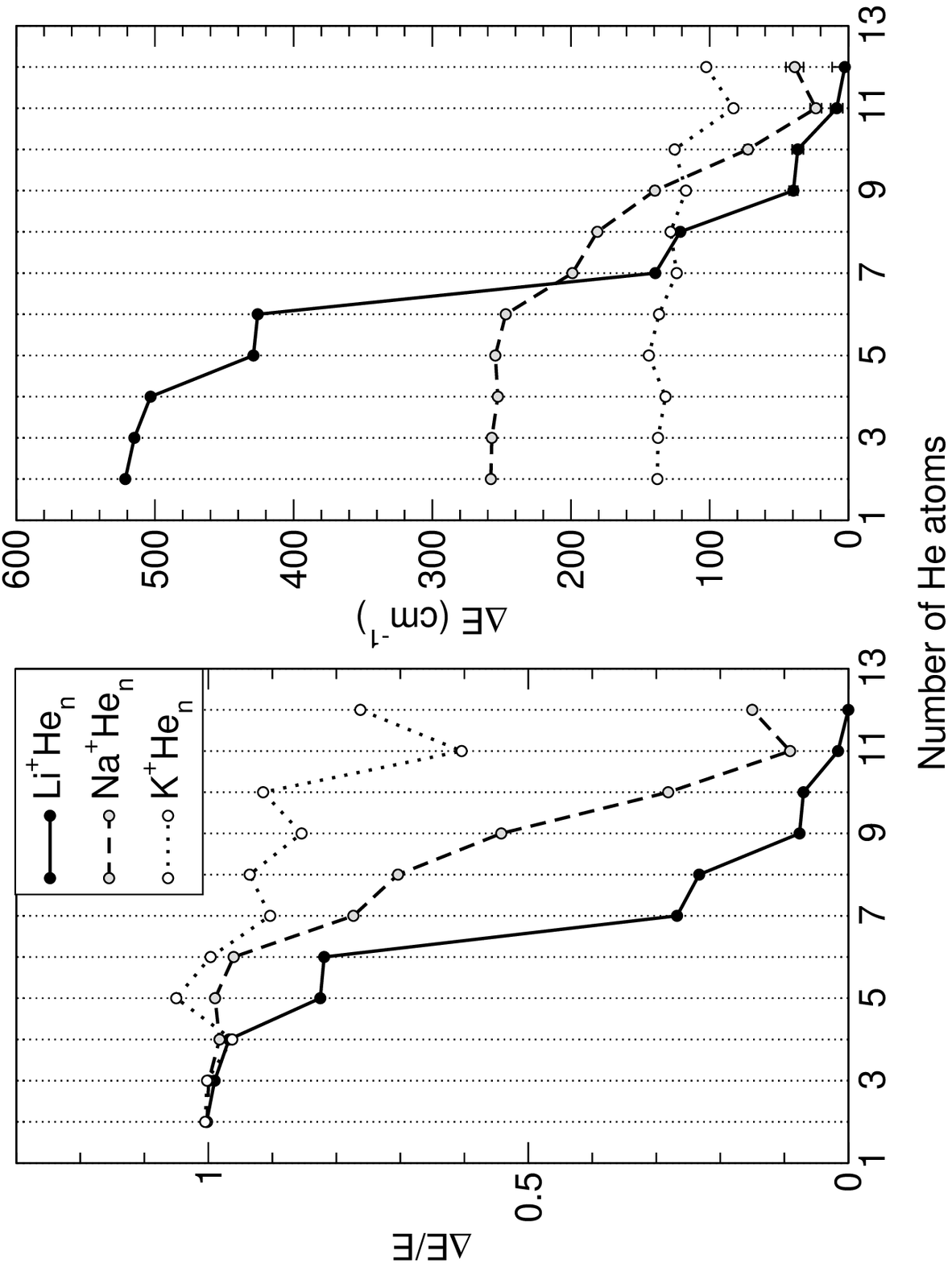}
\label{energies}\caption{}
\end{figure}

In Figure \ref{energies} (right panel) we report the evaporation energies of
the present study up to sizes of $n$=12. As expected with MonteCarlo
simulations our energies are
affected by a statistical uncertainty which, however, is very small compared to
the total energy values. This reduction is certainly due to the prior VMC
optimization procedure. Such
statistical errors are in fact barely visible in
the right panel of Figure \ref{energies}.

The evaporation energy reported by the right panel of Figure \ref{energies} is
defined as $\Delta E=-[E(\mathrm{M^+He_n})-E(\mathrm{M^+He_{n-1}})]$ and
represents the energy  required to evaporate one He atom from the cluster. For
the
Li$^+$ ion we can count two jumps at $n$=6 and 8.  Each jump
reflects clearly the presence of a particularly stable structure (or magic
number). However, as
we shall see below only the flattening at $n$=10 marks the closure of the first
solvation shell (in agreement with our earlier results in \cite{zb:bodo05-e}).
An entirely different situation pertains to the Na$^+$ ion where the decrease of
evaporation energy is continuous and without any marked sign of irregularities.
The most important change occurs for $n=$11 which may then be a magic number.
However, the first solvation shell is probably being completed between $n$=11 and $n=$12.. The
behavior for K$^+$ is even smoother than that of Na$^+$ and does not
present evidence of any particular magic number. The first solvation shell is
likely to be filled at $n=15$ \cite{cl:rossi04}. 

To further point out the energetic behavior of the clusters examined here, we
report in the left panel of Figure \ref{energies} the ''normalized``
evaporation energy i.e. the evaporation energy divided by the dimer energy
($D_0$). The energy behavior of the three species
of clusters is now independent of the strength of the particular Ion-He pair
interaction: up to $n$=6, the behavior of the three clusters is thus
seen to be qualitatively similar.
We can say that the assembly of the cluster's structure is therefore chiefly
driven by
the ion-He pair potential and results in a regime with nearly additive binding
energies. Geometrically we can see
that the cluster  grows by simply adding around the different ions the next He
atom  at a given radial distance which is similar to that identified in Table II. 
The Li$^+$ situation actually presents a
slightly different behavior with respect to the
other two species: the dimension of the central ion is so small that the
crowding of the He atoms and therefore the repulsive He-He interaction in $n$=4 
becomes already important in determining the structure
of the cluster: the addition of one adatom destabilizes the entire cluster
while for Na$^+$ and K$^+$, on the other hand,
the size for which the He-He repulsive interaction becomes
crucially important is $n=6$. 
Furthermore, for $n>6$ the specific nature of the ion is
still important in determining the energetic behavior of the cluster growth: the
bigger the ion, the larger the ratio $\Delta E /E$, which means that for
Na$^+$ and K$^+$ one has a reduced crowding of  He atoms around the dopant
center, hence an overall smaller contribution from the He-He repulsive interactions. 

\subsection{The solvation process: electrostriction effects}
It is interesting now to look at the effect that the presence of a ionic dopant
has on the surrounding distribution of He atoms. As we shall show in this
section,
the net effect of a positive charge is that of a substantial increase of the
density  of He atoms surrounding the latter. This density can get to be so
large that
the He-He pair distribution penetrates well into the repulsive region of the
He-He interaction (electrostriction effect). 

\begin{figure}
 \includegraphics[height=0.9\textheight]{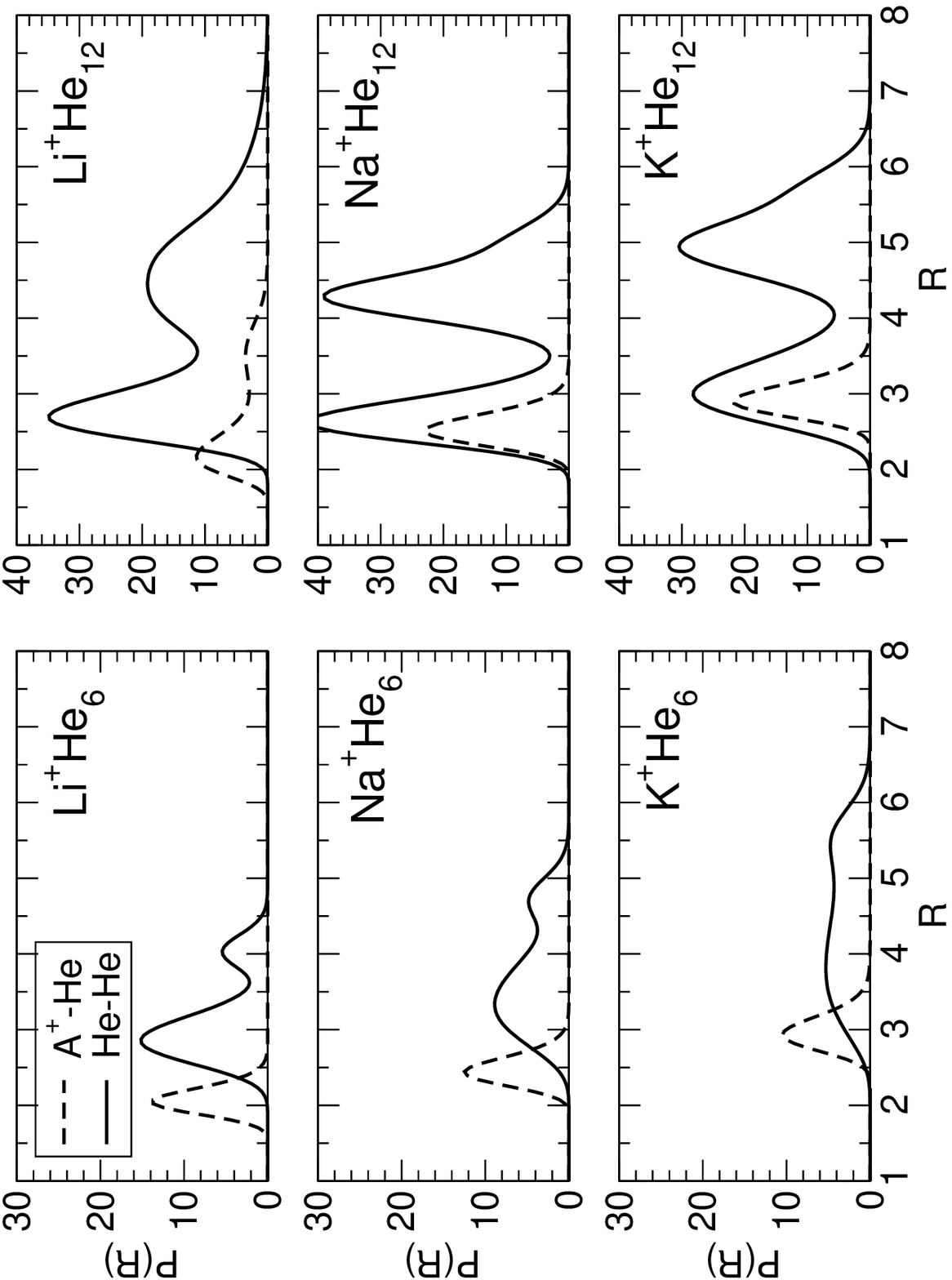}
\label{distr}\caption{}
\end{figure}

\begin{figure}
 \includegraphics[height=0.9\textheight]{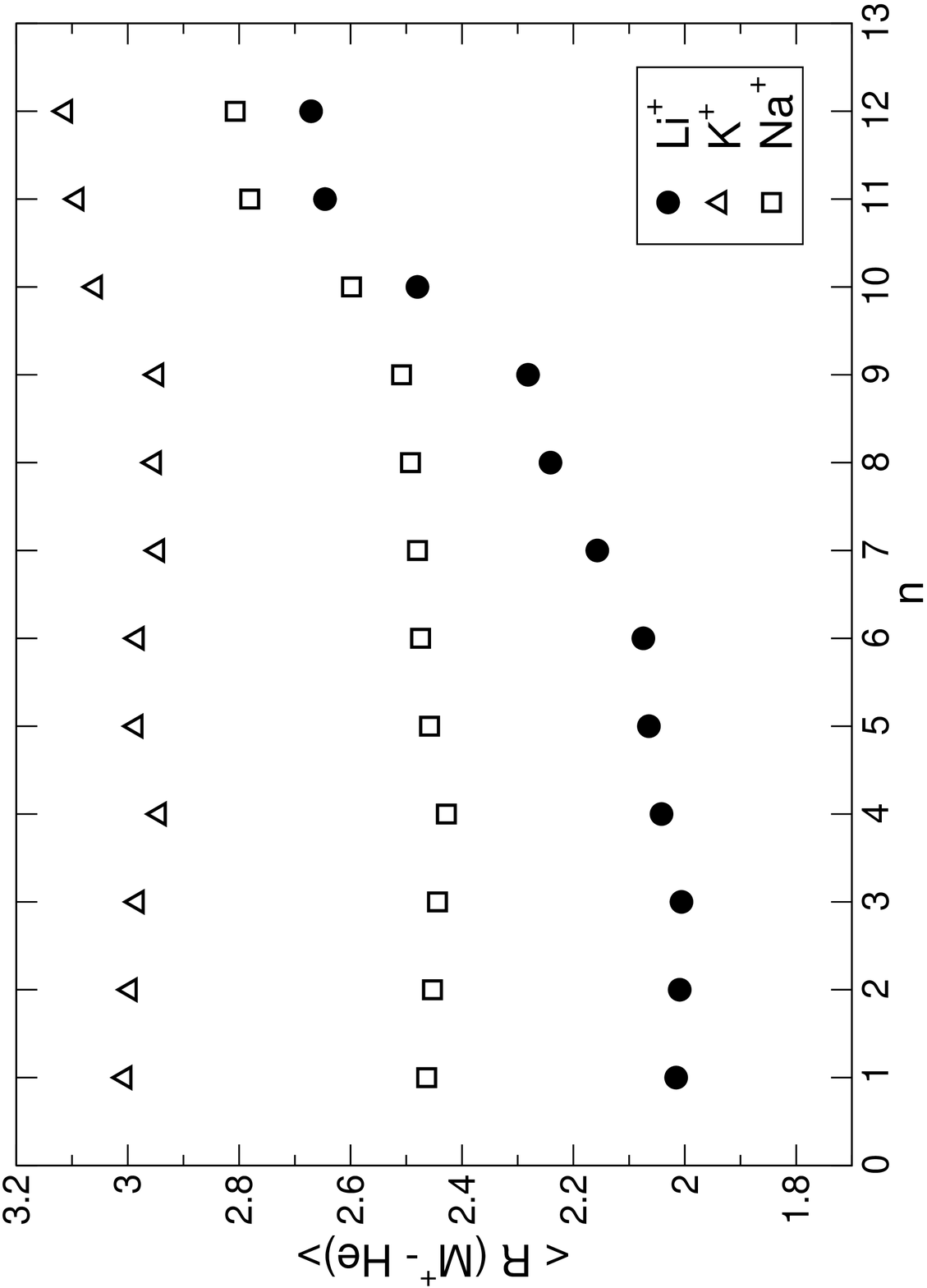}
\label{distance}\caption{}
\end{figure}

In order to prove the above point, the DMC calculation  shown by Figure
\ref{distr} report the Li$^+$-He radial
density distributions  together  with those relative to the He-He distance.
The cluster we present in that figure have
been chosen in order to show the different regimes of solvatation undergone by
the central
ion: up to $n=6$ we clearly see that the average radial position of the He
atoms with respect to the ion (solid lines) is the same, i.e. the 6 He atoms are
part of the first shell. We can clearly see in the Li$^+$He$_{12}$ radial shape  the appearance of a
shoulder in the Ion-He radial distribution which  denotes  the fact that the He
atom positions are not equivalent anymore. On the other hand, the first
solvation shells of Na$^+$ and K$^+$ are larger than for Li$^+$ and therefore we cannot
yet see this effect in their distributions. To further test this point we can look at the average
distances  that we get for the Ion-He pairs inside the various clusters and
that are reported by Figure \ref{distance}. For the Li$^+$ case, we clearly see
a significant increase in the Ion-He distance  after $n=6$ which is mainly due
to a steric effect: on one hand the He atoms tend to assume positions which
maximize the attractive branch of the He-He interaction, on the other hand they tend to reside at the
shortest possible distance from the central ion, thereby minimizing the ion-He
interaction: the increase in distance is
a result of the energy balance between these two opposite driving forces. For the
Li$^+$ clusters we further see an increase after $n=9$  in the radial
distance because the latter is now given by the average between the first and
second shells. A similar behavior can also be noticed also for Na$^+$.

The data reported in Figure \ref{distr} for the He-He distributions (solid lines) show
clearly the existence of two He-He distances. For the Li$^+$He$_6$ cluster the
He-He distribution is typical of an octahedral structure  where we have a
short distance between adjacent helium atoms and a larger one between the apical
ones (see also the similar analysis carried out in
\cite{zb:bodo05-e,zb:bodo06}). From
the fact that the two peaks in the distribution are well separated we can say
that the helium atoms are fairly ''stiff`` also in their relative motion. The
same behavior can be found for larger structures up to $n=12$ where, however
we see that the appearance of an additional peak makes the distribution more
blurred: a new ''shell`` is being formed and the snowball is not yet completed. 

If we look at the He-He distributions for Na$^+$ and K$^+$ clusters we see
clearly how the species with $n=6$ are not as yet characterized by a regular
structure because those clusters are not octahedral ''solids``, as we
shall  further show below.
A localization effect can be noticed, however, for the cluster containing
$n=12$ atoms which has a regular structure (icosahedral) for both ions.

\begin{figure}
 \includegraphics[height=0.9\textheight]{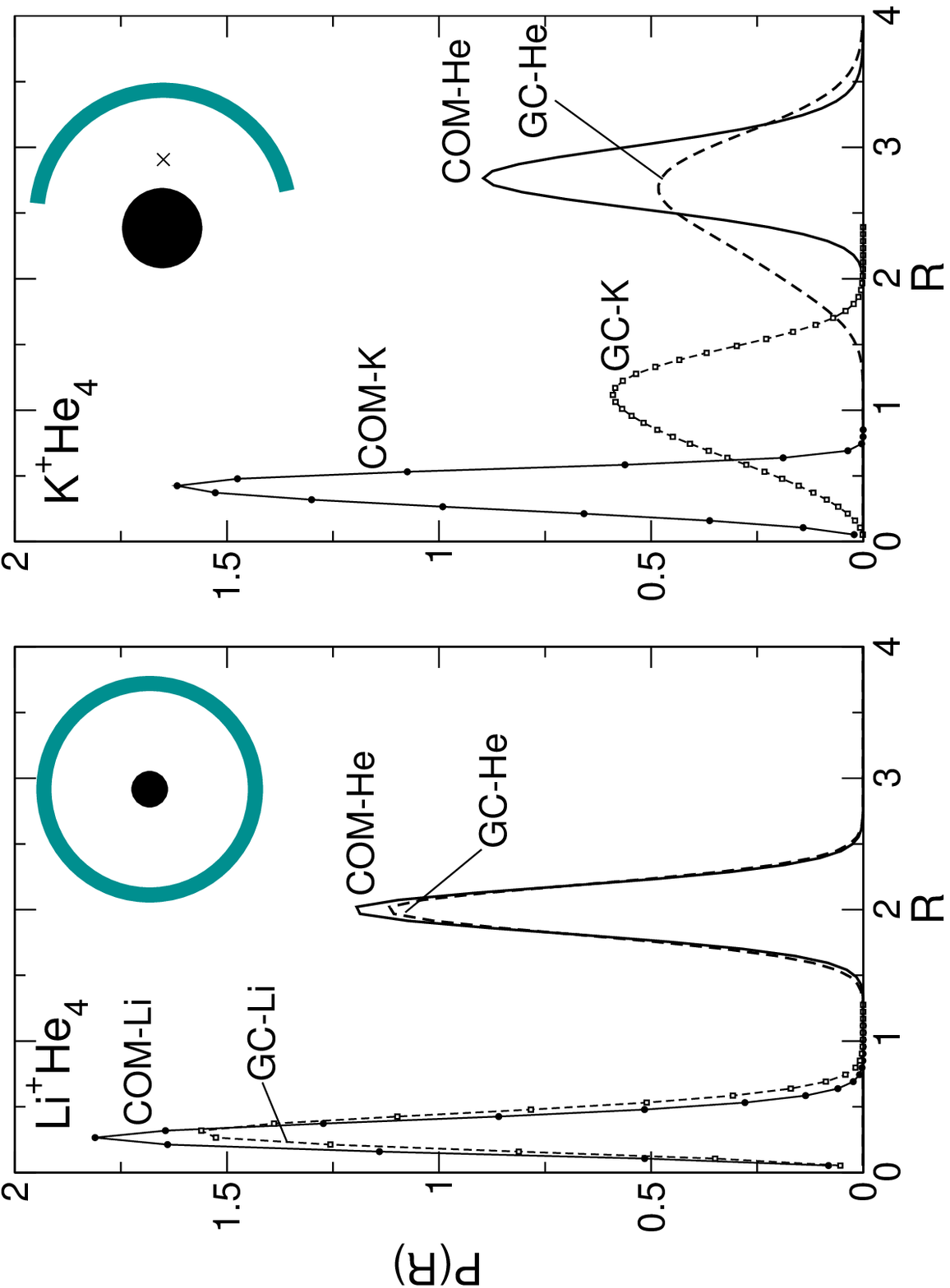}
\label{clu-4}\caption{}
\end{figure}

Putting together the information gathered from the analysis of the
quantum distributions we can say that due to its 
small steric effect, the accretion process of the first solvation shell in
Li$^+$He$_n$ is isotropic, while the same process is not isotropic for the
Na$^+$ and K$^+$ ions which are slightly larger. For Li$^+$, in fact, the
structures with $n \ge 4$ are
fairly compact and therefore the addition of an helium atom has the effect of
increasing the radius of the solvation cage to accommodate it, while however
still remaining
structurally compact. The accretion of the
solvation shell is therefore chiefly dominated by the strong Li$^+$-He
potential. For the other two ions, instead, the solvation process is anisotropic
due to the large size of the ions which does not allow for the creation of
compact structures: only for $n=9$ we see the onset of a nearly complete 
''spherical`` solvation
shell which is marked by an increased in localization of the He-He distributions
(see figure \ref{distr}). In order to further prove this point, we can look
at the geometric distributions which locate the distances of the ion and of the
He atoms from the center of mass (COM) and from the geometrical center (GC, i.e.
the simple geometrical average of the positions) of the
clusters (for more specific definitions, see ref. \cite{cl:dipaola05}). In a
cluster with an isotropic distribution the COM
and the GC of the entire system obviously coincide, while they have different
values for an anisotropic distribution of He atoms especially when the ion
is heavier than He. In Figure \ref{clu-4} we see that
for the small clusters with $n=4$ the solvation is clearly isotropic for Li$^+$
and anisotropic for K$^+$. In other words in the specific case of $n=4$, this
means that the K$^+$He$_4$ is not a tetrahedron, but (in a classical
stick-and-ball
model) the He atoms tend to group on one side of the ion forming an open
''cup``. For larger cluster this anisotropic effect obviously decreases and for
the largest structures that we have analyzed here ($n=12$) it has almost
completely
disappeared.

\begin{figure}
 \includegraphics[height=0.9\textheight]{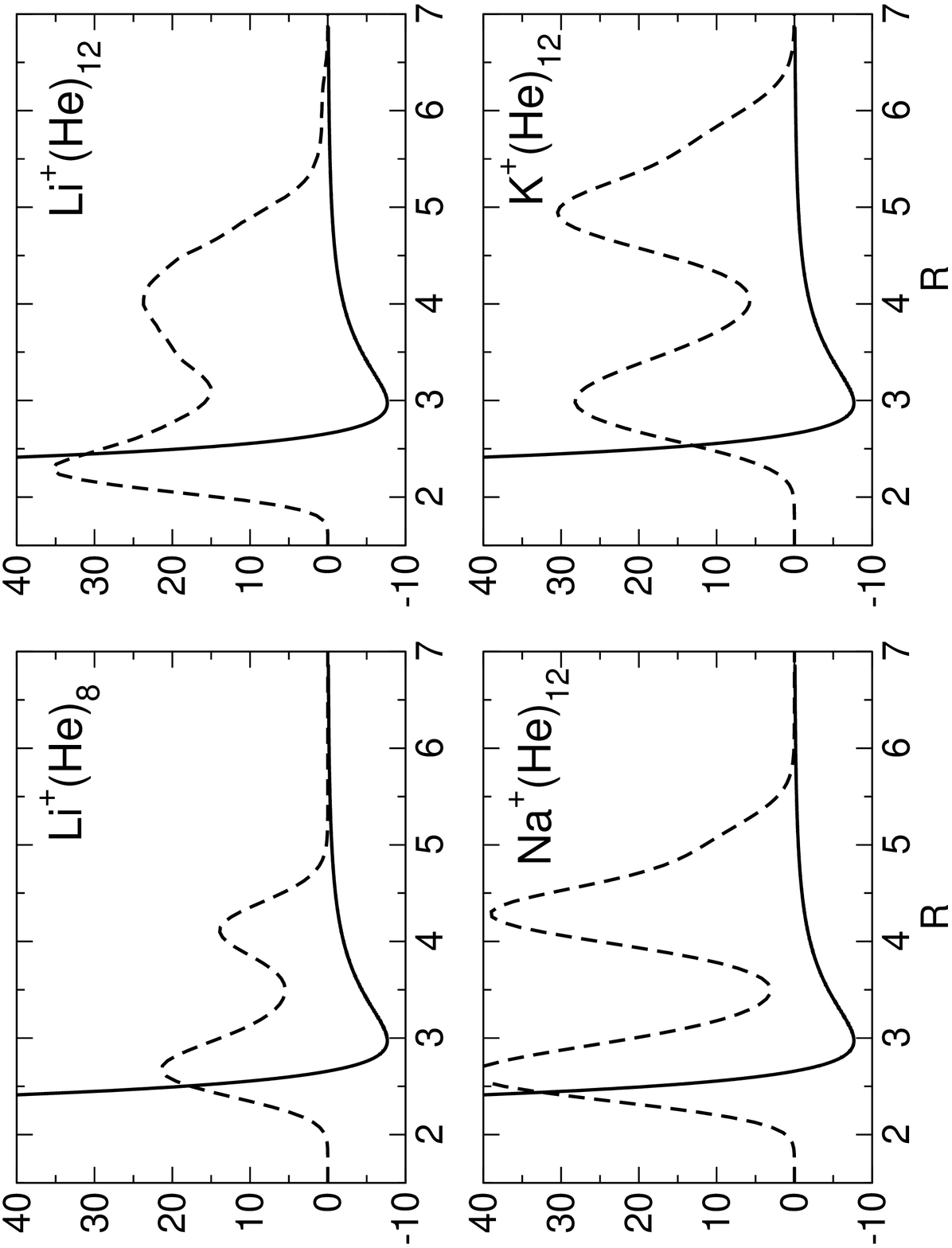}
\label{strict}\caption{}
\end{figure}

As mentioned at the beginning of this work, ''electrostriction`` effects
describe an increase in the density of the quantum
solvent around the cationic impurity with respect to the outer region of the
cluster: to show precisely this feature, 
we report in the upper panels of figure
\ref{strict} the DMC distribution of the He-He radial density superimposed on
the He-He potential. One clearly sees there how, as $n$ increases (see also the
distributions reported in figure \ref{distr}) the He atoms get further drawn
around the ion and therefore their densities show additional
short-range accumulation which could be described as ''tunneling`` across the
repulsive barrier: the destabilizing effect of entering the two body PEC
repulsive region is balanced by the stabilizing contributions coming from the
attractive effects of the Li$^+$-He potential. 

In contrast with the above findings, the lower panels of figure \ref{strict}
shows the  behavior for the same cluster sizes for for Na$^+$ and K$^+$. We
clearly see there that the tunneling is somewhat reduced and
therefore less significant ''crowding`` of the solvent adatoms occurs around
the charge: the electrostriction effects, although still present, are less
marked and therefore less likely to induce  magic numbers in cluster
growth as markedly as in the case of the Li$^+$ cation.

\subsection{The snowball effect}
Another important feature which characterizes the ionic dopants in He droplets
is the presence of a more structured solvent environment around the
ion \cite{cl:galli01,cl:rossi04}:  positive ions are thus surrounded by
many He atoms that are strongly compressed as a result of electrostriction.
The resulting ionic core is thought to be as a solid, with a diameter of several
\AA~ containing many He atoms, and is referred to as a ''snowball``. 
To investigate the possible existence and the geometric structure of such
snowballs we have mapped the
position of the He atoms surrounding the central ion in terms of polar angles
defined via a convenient spherical coordinates system \cite{cl:galli01}. The
system we use here defines a $z$ axis joining the impurity and one He atom
and the $xz$  plane as the plane that contains this $z$ axis and a second He
atom: all the remaining $(n-2)$ He atoms are  mapped in terms of their polar
angles $\theta$ and $\phi$ with respect to that reference frame. We can
therefore generate an angular map of the positions of the He atoms surrounding
the central charge. We focus here on the results that can be obtained for
the most regular structures that we have found in our exploration of
clusters with $n\le12$. 

\begin{figure}
 \includegraphics[angle=-90,width=0.6\textwidth]{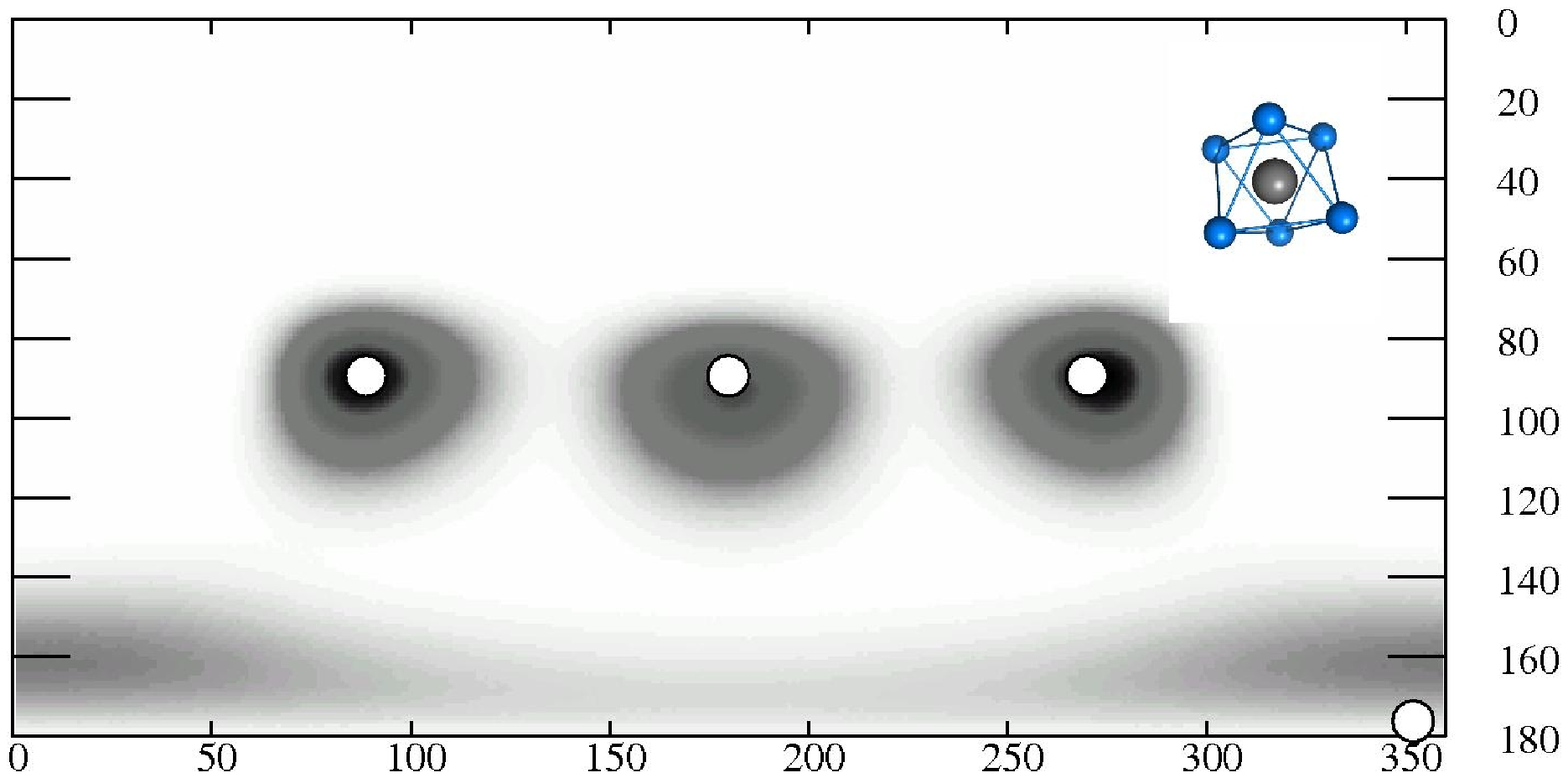}
\label{snow-li}\caption{}
\end{figure}

The Li$^+$ case has a clearly identifiable magic number for
$n=6$ (although probably the first solvation shell is filled by $n=10$).
Although Na$^+$ has no evident magic numbers, from the previous discussion 
about  He-He distributions we expect a special structure for $n=12$.
Also for K$^+$ we present the angular analysis for $n=12$ although the
completion of
the first solvation shell would probably occur at $n=15$
\cite{cl:rossi04,zb:bodo06}. In figure
\ref{snow-li} we can see the angular map of the quantum distribution 
and a ''classical`` reference map that is obtained by simply marking the
angular coordinates of a regular octahedral structure obtained by a
minimization of the corresponding total potential \cite{zb:bodo06}.
When one chooses a rigid octahedral shape, the interpretation of these maps is
very
simple: the $z$ axis goes through one of the vertices of the octahedron so that
we can find the He atoms only at two $\theta$ values: 90$^\circ$ and
180$^\circ$. By fixing the
$xz$ plane on a second vertex, we find the remaining three He atoms at the
$\phi$ values 90$^\circ$, 180$^\circ$ and 270$^\circ$. The quantum distribution
in  figure \ref{snow-li} presents three evident spots at
exactly the positions of the classical rigid structure. The fourth atom, the one
on the second vertex laying on the $z$ axis, has an undefined $\phi$ value, and
is completely delocalized at the bottom of the quantum map where one can see a
weak shading associated with its density distribution. 

The striking similarities between a classical rigid structure and our quantum
distribution is a consequence of the fact that the first solvation shell with
such strongly interacting impurities is  very much like a regular
solid. Because of
electrostriction, in fact, the He atoms are strongly bound and localized near
the central charge. 

\begin{figure}
 \includegraphics[angle=-90,width=0.6\textwidth]{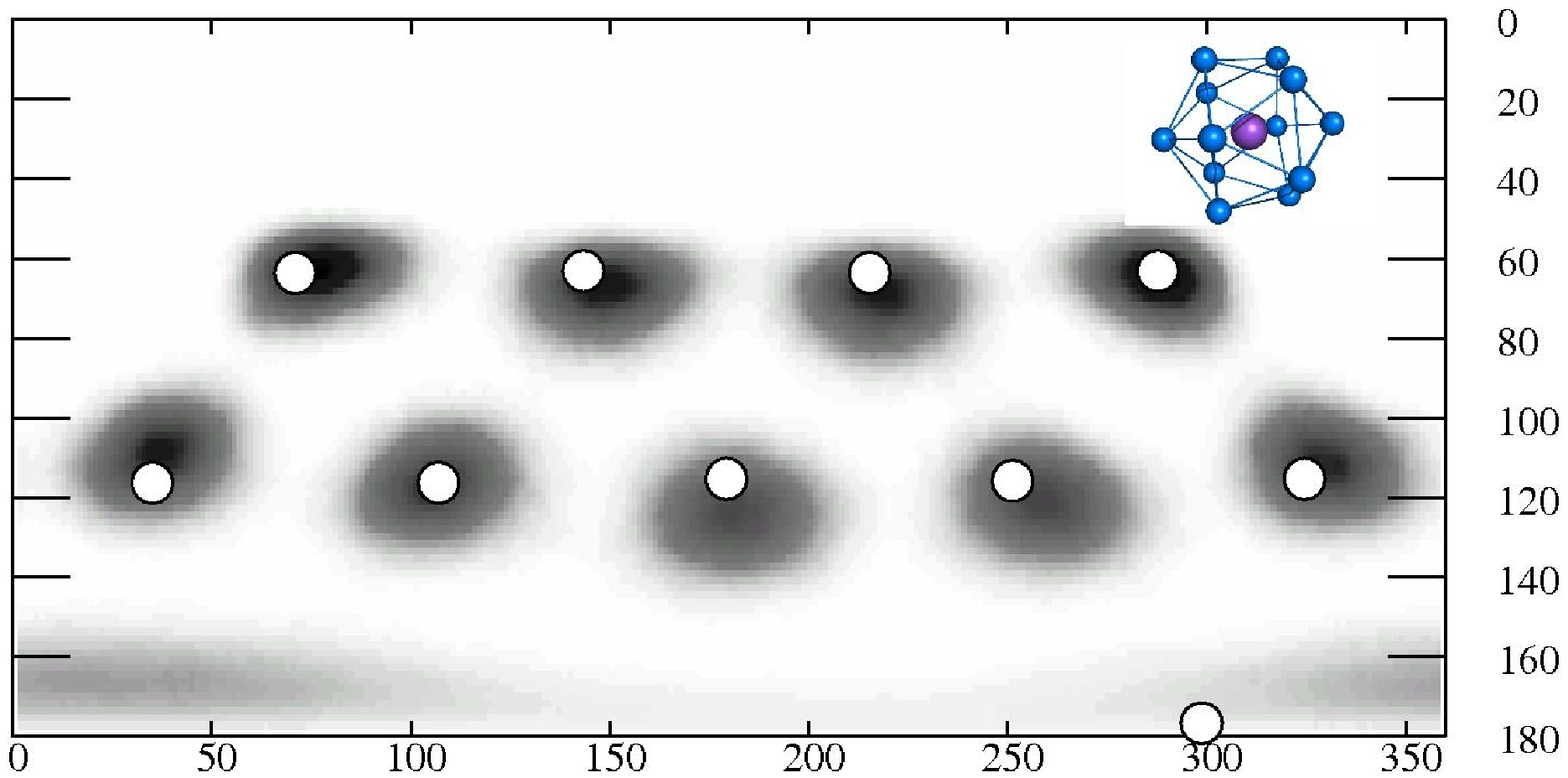}
\label{snow-na}\caption{}
\end{figure}

\begin{figure}
 \includegraphics[angle=-90,width=0.6\textwidth]{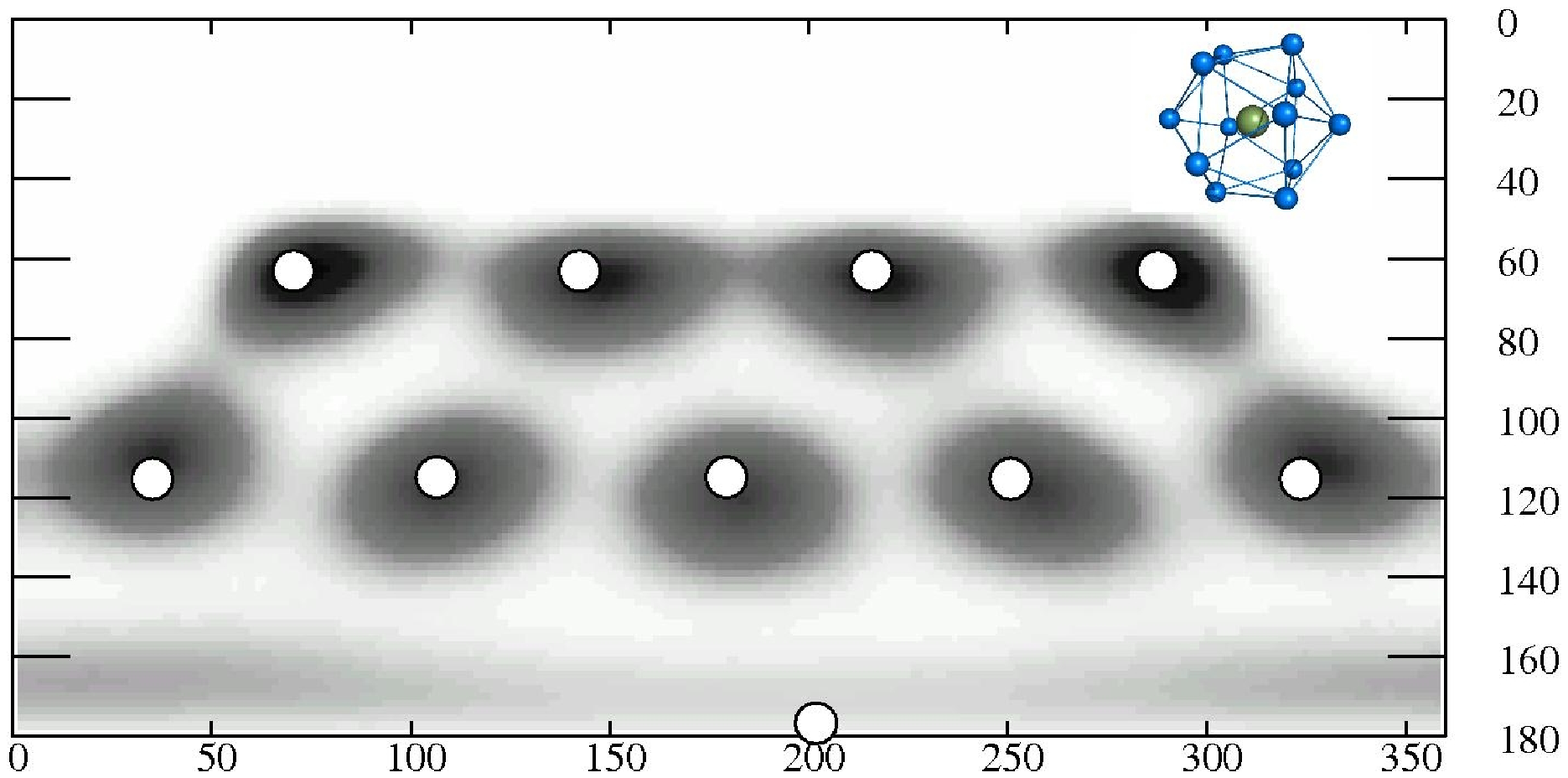}
\label{snow-k}\caption{}
\end{figure}

We have encountered this ''localization'' effect also in the larger cluster for
Na$^+$ and K$^+$. For these ions the most regular structure is the icosahedron
made of 12 He atoms. We report in figures \ref{snow-na} and \ref{snow-k}
the structures for  Na$^+$He$_{12}$ and K$^+$He$_{12}$. Apart from the choice
of the reference system, the situation is completely analogous to what we have
seen for Li$^+$: the quantum structure has a solid-like shape where the He
atoms are localized (less localized for K$^+$ due to its weaker potential) at
the tips of a highly regular solid icosahedra.	This is the first time, to our
knowledge, that such regular structures are so clearly recovered from fully
quantum calculations.

%----------------------------------------------------------------------------
\section{Present Conclusions}\label{last}

In the present work we have studied in detail the microsolvation process of
three similar cations Li$^+$, Na$^+$ and K$^+$ in a quantum solvent, $^4$He. We
have employed a highly accurate quantum method which involves a combination of 
Variational and Diffusion Monte Carlo procedure, amply discussed in the current
literature. The main approximation in the work we
have presented here comes from modelling the total interaction in the
clusters via a sum-of-potentials approximation which could
lead to ``unphysical'' predictions for the very small moieties like 
Li$^+$He$_2$: we have however shown  that the inclusion of 3-Body effects in
the potential seems not to alter the situation for Na$^+$ and K$^+$ and for
all the clusters with $n<4$.

We have also focussed on  the
accretion process of the very first solvation shells around the dopant ion in
order to establish  geometric structures and the presence of some
structural regularity in them. All the calculations have included clusters up
to $n=12$ He atoms. In all the three cases the impurity is placed inside
the cluster and acts as a coordination center to which the solvent atoms are
bound. The radial distributions of the Ion-He distance show invariably a very
high degree of localization which is mainly due to the strong Ion-He potential
which dominates over the weaker He-He interaction, at least for the smaller
structures. Only for the smallest ion, Li$^+$, we have seen a change of this
behavior when passing to $n>10$ where the distribution becomes clearly made of
two merged peaks, the second one representing the onset of the second solvation
shell, i.e.
atoms which are invariably localized at a larger distance from the ion. 

The He-He distribution, on the other hand, shows a much higher degree of 
dispersion especially for K$^+$. This fact can be explained by considering that
the K$^+$-He interaction is the weakest of the three PECs and that the 
dimension of this ion is also the
biggest: the He atoms are far from the coordination center, their motions can
span a much larger volume than in the case of Li$^+$ and therefore
they turn out to be more delocalized. 
This effect somehow counteracts the
electrostriction effect which consists in the increase of local He density in
the proximity of the ion. Indeed we have seen how in the case of a small ion
like Li$^+$ the He-He distribution clearly shows this effect by forcing the He
atoms to penetrate considerably into the classically forbidden region of the He-He
pair potential. This effect is still present, although less marked, for Na$^+$
and tends to be largely absent for K$^+$ unless one considers the bigger 
regular structures like the one with $n=12$. 

Due to this strong electrostriction effect, the Li$^+$ clusters present 
several magic numbers that are clearly marked by a decrease of the evaporation
energy. For these clusters the solvation process remain isotropic along all the
series up to $n=10$ where we have seen the filling of the first shell.

On the other hand, the magic numbers
are less identifiable in the Na$^+$ clusters although we still see some
structure in the  dependence of the evaporation energy on cluster size. We have also 
computational evidence that
the highly regular structure with $n=12$ marks the closure of the first
solvation shell, but more calculations would be required to prove it. It is
interesting to note
that contrary to Li$^+$, for Na$^+$ ions the growth of the first solvation
shell is not isotropic but forms a kind of ``cup'' on one side of the ion as
we have showed using the quantum  radial distribution with respect to the
geometric center. The situation is similar for K$^+$ where we further see no
trace of magic numbers.

The shape of the quantum distributions of the angular coordinates between 
various atoms in
the clusters allows for the presence of  regular structures which show a
striking similarities with
their classical counterparts and a surprisingly high degree of angular
localization for the solvent atoms: one clearly sees very ordered structures
typical of  solid-like geometries. This quantum analysis confirms the
qualitative validity of a the
snowball model although points out that the snowball structures, 
as well as their spatial extension, strongly depend on the chemical nature of
the dopant ion.

\begin{acknowledgments}
Financial support from the EU Network COLMOL HPRN-CT-2002-00290, from the 
Scientific Committee of the University of
Rome, from the Ministry of University and Research (MIUR) Nationa Projects
(PRIN) and from the CASPUR Supercomputing Center are all gratefully
acknowledged. We also thank the Agnelli
Foundation for financing visits between Rome and Ko\'c universities
during the development of the present research.
\end{acknowledgments}

\newpage
\section* {caption to figures}
\begin{itemize}
\item{Fig 1: DMC radial distribution of the Ion-He and He-He distances in
the $n=2$ cluster}

\item{Fig 2: Right: Evaporation energy
$\Delta E=-[E(\mathrm{M^+He_n})-E(\mathrm{M^+He_{n-1}})]$. Left:
Dimensionless $\Delta E / E $ evaporation function}

\item{Fig 3: Radial A$^+$-He and He-He DMC distribution for A$^+$He$_n$. Distances in
\AA. Ditributions in \AA$^{-1}$}

\item{Fig 4: Average distance between the ion and the He atoms in the three
series of clusters as a function of size. Distances are in \AA.}

\item{Fig 5: Radial distribution of the distance from the center
of mass and the geometrical center of the cluster. Left panel: Li$^+$, right
one: K$^+$. A schematic view of the
cluster geometry is also depicted: in the case of K$^+$ the GC
is indicated by a cross.}

\item{Fig 6: Radial He-He DMC distribution for A$^+$He$_n$
superimposed to the He-He potential. Distances in \AA~ and
energy in cm$^{-1}$.}

\item{Fig 7: Angular correlation maps for Li$^+$He$_6$. $\theta$ is
on the $y$ axis and $\phi$ on the $x$ one. The white dots are the position of the corresponding 
localized minimum structure which is also reported in the small inset.}

\item{Fig 8: As in figure 7 but  for Na$^+$He$_{12}$.}

\item{Fig 9: As in figure 7 but  for K$^+$He$_{12}$.}

\end{itemize}

\newpage
\pagestyle{empty}

\newpage
\pagestyle{empty}

\newpage
\pagestyle{empty}

\newpage
\pagestyle{empty}

\newpage
\pagestyle{empty}

\newpage
\pagestyle{empty}

\newpage
\pagestyle{empty}

\newpage
\pagestyle{empty}

\newpage
\pagestyle{empty}

\end{document}